\documentclass[sigconf,natbib]{acmart}

\usepackage{todonotes}
\usepackage{booktabs}
\usepackage{titlecaps}

\usepackage{multirow}

\newcommand{\up}{$\blacktriangle$}
\newcommand{\down}{$\blacktriangledown$}
\newcommand{\base}{$\star$}

\newcommand{\bm}[0]{BM25}
\newcommand{\titleq}[0]{Query}
\newcommand{\allq}[0]{Topic-Expansion}
\newcommand{\sectionq}[0]{Topic-Aggregation}
\newcommand{\bmTitle}[0]{\bm{}-\titleq{}}
\newcommand{\bmAll}[0]{\bm{}-\allq{}}
\newcommand{\bmSection}[0]{\bm{}-\sectionq{}}
\newcommand{\leadq}[0]{Description}
\newcommand{\meanq}[0]{Mean}
\newcommand{\qqqsm}[0]{Q3SM}
\newcommand{\qqqsmLead}[0]{\qqqsm{}-\leadq{}}
\newcommand{\qqqsmMean}[0]{\qqqsm{}-\meanq{}}
\newcommand{\qqqsmTitle}[0]{\qqqsm{}-\titleq{}}
\newcommand{\sbert}[0]{SBERT}
\newcommand{\cosine}[0]{Cosine}
\newcommand{\euclid}[0]{Euclid}
\newcommand{\sbertCosine}[0]{\sbert{}-\cosine{}}
\newcommand{\sbertEuclid}[0]{\sbert{}-\euclid{}}
\newcommand{\TLouvain}[0]{T5-Louvain}
\newcommand{\TRedundancy}[0]{T5 redundant}

\newcommand{\summary}[0]{summary}

\usepackage {xcolor}
\newcommand{\ld}[1]{\textbf{\textcolor{red}{TODO LD: #1}}}
\renewcommand{\ld}[1]{}

\newcommand{\cut}[1]{}

\newcommand{\progress}[1]{\smallskip \textcolor{blue}{\hfill{} #1 \hfill{}} \smallskip}
\renewcommand{\progress}[1]{}

\renewcommand{\paragraph}[1]{\smallskip  \textbf{#1.}\hspace{1ex} }

\AtBeginDocument{%
  \providecommand\BibTeX{{%
    \normalfont B\kern-0.5em{\scshape i\kern-0.25em b}\kern-0.8em\TeX}}}

\acmConference[arXiv]{}{2023}{Internet}
\acmBooktitle{arXiv} 
\acmYear{2023}

\begin{document}

% \title{A More Realistic Environment for\\Multi-Document Summarization}
%\title{Query-specific Retrieve, Cluster, Summarize:\\A Joint Baseline System and Evaluation for Article Generation}

\title{Retrieve-Cluster-Summarize: An Alternative to End-to-End Training for Query-specific Article Generation}

\author{Connor Lennox}
\email{connor.lennox@unh.edu}
\affiliation{%
  \institution{University of New Hampshire}
  \city{Durham}
  \state{New Hampshire}
  \country{USA}
  \postcode{03824}
}

\author{Sumanta Kashyapi}
\email{Sumanta.Kashyapi@unh.edu}
\affiliation{%
  \institution{University of New Hampshire}
  \city{Durham}
  \state{New Hampshire}
  \country{USA}
  \postcode{03824}
}

\author{Laura Dietz}
\email{dietz@cs.unh.edu}
\affiliation{%
  \institution{University of New Hampshire}
  \city{Durham}
  \state{New Hampshire}
  \country{USA}
  \postcode{03824}
}

% \author{No Author Given}
% \affiliation{
%     \institution{No Institute Given}
%     \country{}
% }

% \renewcommand{\shortauthors}{Lennox et al.}
\renewcommand{\shortauthors}{No Author Given}

\newcommand\blfootnote[1]{%
  \begingroup
  \renewcommand\thefootnote{}\footnote{#1}%
  \addtocounter{footnote}{-1}%
  \endgroup
}

\begin{abstract}
    Query-specific article generation is the task of, given a search query, generate a single article that gives an overview of the topic. We envision such articles as an alternative to presenting a ranking of search results. While generative Large Language Models (LLMs) like chatGPT also address this task, they are known to hallucinate new information, their models are secret, hard to analyze and control. Some generative LLMs provide supporting references, yet these are often unrelated to the generated content. As an alternative, we propose to study article generation systems that integrate document retrieval, query-specific clustering, and summarization. By design, such models can provide actual citations as provenance for their generated text. In particular, we contribute an evaluation framework that allows to separately trains and evaluate each of these three components before combining them into one system. We experimentally demonstrate that a system comprised of the best-performing individual components also obtains the best F-1 overall system quality.

\end{abstract}



\maketitle

% \blfootnote{Code and data will be released upon acceptance.}

% \vspace{-2em}

\section{Introduction}
\label{sec:rqs}
\label{sec:introduction}

Information Retrieval tasks usually stop at producing a ranking of relevant results. In this paper, in line with the SWIRL II report \cite{allan2012frontiers}, we imagine to use these search results and instead present the user with a longer article, that covers all of the relevant \emph{information}. This benefits the users who seek a general understanding of  topics, without having to sift through multiple web pages with redundant content. This would also support users of spoken interfaces or small screen devices, who are not well-served with a ranking.

%%% Task: article generation (and why helpful)
\textbf{Query-driven Article Generation Task:} Given a broad topical search query, generate a relevant article (i.e., a long-form summary).  
The ideal article should contain only relevant information, avoid switching back and forth between topics, while accurately representing the content.

To generate such a query-specific article, search results could be retrieved, organized into relevant subtopics, then condensed with summarization.  We anticipate that the quality of the resulting article will depend on getting all these steps right.
Each of these tasks---retrieval, clustering, summarization---is usually studied in different communities, each using their own benchmarks. Joint inference models and end-to-end training gained popularity when it was shown that systems with individually trained components would suffer from error-propagation which often leads to diminished results. Unfortunately, end-to-end training is often too expensive as greater task complexity requires disproportionally more epochs to converge.  In this paper we explore a cheaper alternative: We demonstrate that system components can be trained individually when we are smart about how to create task-specific benchmarks that are agreeing with one another. 
%using the Wikimarks approach \cite{dietz2022wikimarks}, which automatically harvests test collections for passage retrieval, query-specific clustering, and summarization from Wikipedia. 
 
% \begin{figure}
% \ld{candidate for cutting}
%     \begin{verse}
        
%     \textbf{Query: Natural Resources}

% Natural resources are classified primarily by their...

% Resource extraction is restricted in areas near...

% In recent years, resource depletion has become...
%     \end{verse}
%     \caption{Example article generated in response to a query.}
%     \label{fig:example}
% \end{figure}

\newcommand{\rqWikimarksEvaluation}{RQ1}
\newcommand{\rqGoodBaselineSystem}{RQ2}
\newcommand{\rqManualSummarization}{RQ3}

We contribute (1) experiments with this new benchmark creation and evaluation paradigm, (2) use the benchmark to explore various choices for each component, and (3) seek answers to the following research questions:

\begin{itemize}
    \item[\rqWikimarksEvaluation{}:] Does the benchmark creation approach enable component-wise training for obtaining good overall system performance?
    \item[\rqGoodBaselineSystem{}:] Which baseline system  offers good performance?
    \item[\rqManualSummarization{}:] How does the retrieve-cluster-summarize baseline compare to manually selected ideal passages?
\end{itemize}

% \begin{itemize}
%     \item \textbf{RQ1:} What impacts do non-ideal inputs have on multi-document summarization techniques?
%     \item \textbf{RQ2:} How does the relative change in summarization methods correlate with the performance of upstream retrieval and clustering components?
%     \item \textbf{RQ3:} Which combination of upstream components results in the best performing article generation system?
% \end{itemize}

% \paragraph{Outline} In Section \ref{sec:related_work} we discuss related work. In Section \ref{sec:approach} we discuss the joint system as well as choices for its retrieval, cluster, and summarization components. In Section \ref{sec:eval} we describe the evaluation of the system and discuss results in Section \ref{sec:results} before concluding the paper.

\progress{End of Page 1}

\section{Related Work}
\label{sec:related_work}

\paragraph{Query-specific clustering and metric learning}
Clustering and topic modeling aim to find coherent subgroups within a set of documents \cite{blei2003latent}. An effective approach is to combine Hierarchical Agglomerative Clustering with metric learning. The similarity metric can be learned with neural methods such as Sentence BERT \cite{reimers2019sentence}. For IR tasks, similarity learning and clustering can be improved by taking the query into consideration, as demonstrated by \citet{bernardini2009full}. 
%Using the original search query as an input to a clustering system allows for clusters to be targeted towards relevant subtopics of the query .
 Kashyapi et al.\ \cite{kashyapi2022query} improve Sentence BERT by incorporating the query into the metric learning.

\paragraph{Integrating Retrieval with Clustering and Summarization}
Within the IR context, clustering is studied in order to organize retrieved results into coherent groups that emphasize nuanced subtopics \cite{carpineto2012consensus}. Clustering has been used for search result diversification \cite{santos2015search}, and query expansion \cite{carpineto2012survey}.
 \citet{raiber2013ranking} demonstrate how clusters can improve search results.

Retrieval-augmented language models \cite{lewis2020retrieval,izacard2022atlas} integrate retrieval with other NLP tasks, using end-to-end training. Claveau \cite{claveau2021neural} leverages text generation to expand the search query.

% \ld{Unclear: The context of information retrieval also provides the context of a query, leading to a query-specific clustering task} where documents are clustered into subtopics of a given search query \cite{bernardini2009full}.

%\ld{Something about search result diversification}

\paragraph{Query-driven article generation}
Automatically generating informative articles is a problem that has been studied by both learning document templates \cite{sauper2009automatically} and neural text generation with query-focused multi-document summarization \cite{hayashi2021wikiasp}. 

%In addition, there has been significant work in studying retrieval and other information retrieval tasks within the context of Wikipedia \cite{dietz2017trec}.
\citet{banerjee2015wikikreator} seek to expand incomplete Wikipedia pages (called ``stubs'') by leveraging information about a topic found via web retrieval. \citet{liu2018generating} perform a two-phase ``extract then abstract'' approach to improve the generated output. In contrast, we organize relevant text with document retrieval and clustering, then collate the information with summarization.  

\paragraph{Cluster-based summarization}
A common approach is to apply multi-document summarization to retrieved search results \cite{lin2002single}. T5 \cite{raffel2019exploring} and GPT \cite{radford2019language} are commonly used to drive summarization. Training costs can be reduced by limiting token ranges \cite{ma2020multi}, by exploiting  the graph structure \cite{li2020leveraging} or by operating on clusters of documents and using Longformer-based neural models \cite{pasunuru2021efficiently}. %Additionally, we believe that summarization will benefit by making queries and relevant clusters explicit.

\paragraph{ChatGPT and generative LLMs}
The success of large generative language models trained with human feedback \cite{ouyangtraining}, such as ChatGPT, is achieved with a simple neural GPT architecture \cite{radford2019language} and lots of training data. However, there are known issues pertaining to the correctness of provided facts \cite{guo2023close}.  A common issue is their inability to trace the generation back to source information, making fact-checking very difficult. We focus on an alternative bottom-up approach that allows to attribute individual passages to the original source documents.

\progress{halfway down Page 2}

\section{Approach}
\label{sec:approach}

\subsection{Overarching System}
% move to approach
We develop and study a system that given a query $q$ and a corpus of documents, produces an article that covers all relevant information. Our system consists of three components: 
%
%a retrieval system that obtains the top $k$ documents for the given query; a query-specific clustering method that groups these documents into $K$ subtopic clusters; and a summarization method that condenses the documents in each cluster and combines the text to form the article.

\begin{enumerate}
    \item The information retrieval system that retrieves top $k$ documents for the given query.
    \item A query-specific clustering method that groups these documents into $K$ subtopic clusters. 
    \item A summarization component that condenses documents of each cluster and combines the text to form the article.
\end{enumerate}

 To simplify our experiments, we use top $k=50$ documents and during clustering we use $K$ as the number of true subtopics.

\begin{figure}
    \centering
    \includegraphics[width=0.9\columnwidth]{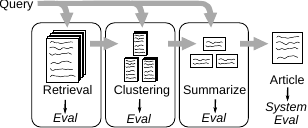}
    \caption{Article generation system. Evaluation of individual components and joint system on the same test collection.}
    \label{fig:approach}
\end{figure}

Given the myriad of retrieval, clustering, and summarization approaches, a central question is how to select the best method for each component and train it. Neural approaches are available for each, so theoretically it is possible to perform end-to-end training. However, the high training costs and slow convergence renders such a model infeasible on GPUs available to academics. 

\subsection{Joint System Evaluation Paradigm}\label{sec:evalparadigm}

Our main argument is that by being smart about how to create task-specific benchmarks, we can avoid error propagation as typically seen with pipelined training approaches and hence provide an alternative to expensive end-to-end training.  
%\ld{drop?} We demonstrate the utility of our benchmark creation approach on a variety of query-specific and query-agnostic approaches for each component listed in Section \ref{sec:components}, and develop a competitive baseline system for future research.

The key to success is to create benchmarks for each sub-task that share a notion of relevance, so that all components work in unison when the system is composed. For our example system this means that (1) documents marked as being from the same cluster are also relevant; while (2) all clusters are represented in the gold summary. We call these \textsl{task-coordinated benchmarks}.

Such benchmarks can be harvested from existing collections of structured articles. The Wikimarks approach \cite{dietz2022wikimarks}  derives benchmarks for different relevance-oriented tasks from Wikipedia articles. Similar ideas have been successfully applied before \cite{hayashi2021wikiasp}, and studies have shown that these strongly agree with manually created relevance annotations \cite{dietz2020humans}.
Our goal is to generate an article from search results that mimics the style of Wikipedia pages, which discuss broad topics. The Wikimarks approach creates a useful training benchmark as follows:
(1) First select Wikipedia pages that correspond to desired information needs. The title of each page is interpreted as a search query, such as \emph{``Natural Resources''} as an example.
(2) Sections of the page are interpreted as relevant subtopics, with the heading representing the name of the subtopic, such as  \emph{``Depletion''}, \emph{``Protection''}, and  \emph{``Classification''} of natural resources. 
(3) Any paragraph on the page is considered relevant. Additionally, each paragraph is associated with a subtopic (or multiple subtopics for section hierarchies). 
(4) The visible text of the Wikipedia page is used as a gold \summary{}.

This paradigm allows us to evaluate (and train) each component separately: The retrieval stage is evaluated by retrieving all relevant paragraphs in response to the search query derived from the page title. The clustering is evaluated by grouping paragraphs according to sections. The summarization component is evaluated by its agreement with the gold \summary{}.

Since all subtasks are trained on benchmarks that naturally agree with one another, we hypothesize that the compound system of individually trained component would also obtain best system-level performance.

\section{Studied System Components}\label{sec:components}

\ld{retrieval-awareness across all components}

To explore the feasibility of our evaluation paradigm  and to identify a reasonable baseline system for future research, we explore the following variety of approaches for the retrieval, cluster, and summarization component. %We focus on approaches with known performance characteristics to showcase the benchmark creation and evaluation paradigm.

\subsection{Retrieval}

While any retrieval method could be used, in this work we focus on methods with well-known performance characteristics, such as well-known traditional retrieval models using BM25 (Lucene, standard parametrization). 

\paragraph{\bmTitle{}} We retrieve a BM25 ranking, which is known to perform moderately well on the benchmark. 

\paragraph{BM25-Topic} To study the effects of strong versus weak retrieval models, we include rankings that include oracle knowledge of relevant subtopics, which are taken from section headings: (1) \textit{BM25-Topic-Expansion} expands the query with keywords of query-relevant subtopics. (2) \textit{BM25-Topic-Aggregation} retrieves search results for each subtopic (using query and topic description), then merges rankings with unsupervised reciprocal rank aggregation $\sum_{\mathcal{R}}\frac{1}{rank_{\mathcal{R}(d)}}$.

\subsection{Clustering}
\label{sec:clustering}

Retrieved documents are clustered with the goal of identifying relevant subtopics pertaining to the original information need. We study two methods which learn a topic-sensitive similarity metric for a pair of documents, then use these similarities via hierarchical agglomerative clustering to produce clusters. %We ask all models to produce $K$ clusters, based on the true number of subtopics (oracle).
We study:% several methods of subtopic clustering:

\paragraph{\sbert{}} We use query-agnostic Sentence-BERT \cite{reimers2019sentence} vector representations of each document. Representations are fine-tuned so that documents are more similar when they are in the same cluster. 
As a similarity, the authors suggest to use \textit{Euclidean} distance, but we also include an experiment with \textit{Cosine} similarity.

\paragraph{\qqqsm{}} To encourage subtopics that are relevant for the query, the Query-specific Siamese Similarity Metric (QS3M) \cite{kashyapi2022query} incorporates similarities between queries and document representations into Sentence-BERT, and is known to outperform Sentence-BERT.

We study three query models for use in \qqqsm{}: (1) \textit{\titleq{}}, which uses the query text; (2) \textit{\leadq{}}, which assumes a description of the query (such as lead text); (3) \textit{\meanq{}}, which uses the centroid of document vectors.

% Additionally, in the context of ideal retrieved documents, we can consider an idealized clustering system. As ideal documents are passages extracted from a ground-truth article, we use top-level article headings to define clusters. This relationship is not possible to extract when we are using documents selected via an information retrieval system, however.

%\vspace{-1em}
\subsection{Summarize}

% \begin{enumerate}
%     \item for each cluster $c$:
%     \begin{enumerate}
%     \item for each document $d$ in the cluster, produce a single-document summary $s_d$.
%     \item identify redundant summary sets $g$, via Louvain clustering of Tf-idf vectors of summaries $s_d$.
%     \item produce one summary $s_g$ for each redundant set $g$, by concatenating $s_d'\in g$ and using single-document summarization.
%     \item order the non-redundant summaries $s_g$ by starting with the largest redundant set, then greedily choosing the next summary $s_{g'}$ by SBERT similarity.
%     \item generate a sub-topic section by aggregating all non-redundant summaries $s_g$ in this order.
%     \end{enumerate}
%     \item append all subtopic sections to generate the article
% \end{enumerate}

%%%%

 The cluster-sensitive summarization model will summarize each cluster into its own section, which are then combined into an article. 
 
 \paragraph{\TLouvain{}} As an underlying summarization method, we use T5 \cite{raffel2019exploring}\footnote{T5 model \url{https://huggingface.co/t5-base}} model\ld{ out-of-the-box}, which has been found to work well for information retrieval tasks \cite{pradeep2021expando}. When applied to our documents, we find that T5 typically produces summaries of one or two sentences.

To produce an article section from a cluster of documents, we first obtain a preliminary summary each document. 
 We the literature and use the Louvain clustering algorithm \cite{blondel2008fast} to detect redundant sets among these summaries and generate a T5 summary for each redundancy-free set. 
 By preserving references to the original search results, our system can provide provenance for each summary. In \textit{\TRedundancy{}} we skip the redundancy removal step.

From each cluster, a corresponding section is generated by arranging redundancy-free summaries $s$: 
These are ordered starting with the summary of the largest redundancy set, then 
%, as it contains information that is frequently mentioned across this subtopic. Then, we 
greedily choosing the most similar summary $s_j$ next, using Sentence BERT's similarity (as trained in Section \ref{sec:clustering}).

\progress{Halfway down page 3}

\section{Evaluation Paradigm}
\label{sec:eval}

We use the following experimental setup to evaluate our system and discuss the research questions raised in Section \ref{sec:rqs}.

We use a Wikimarks \cite{dietz2022wikimarks} dataset for all three components from the TREC-Complex Answer Retrieval \cite{dietz2017trec} dataset. Using both manual and automatic benchmarks, we train components on 117 article-level queries of benchmarkY1train and test on 126 queries of benchmarkY1test (omitting six queries that don't have enough sub-topics). The document corpus consists of 30 million deduplicated paragraphs from all Wikipedia articles. 

The retrieval stage uses the whole corpus as input and is evaluated on the automatic benchmark. The clustering stage is trained to cluster all relevant paragraphs according to the sections of the automatic benchmark. The summarization system is trained to summarize passages from the manual benchmark to generate the corresponding sections of the gold article.

%
%For each combination of components, we use our system to generate an article for each of the
%126 \ld{why are some missing?} 
%queries. 

\begin{table}
\caption{Component-wise results for retrieval in (Mean) Average Prediction and clustering  in Adjusted RAND Index.}
\begin{footnotesize}

\begin{tabular}{ll}
\toprule
Retrieval Method       & MAP             \\ \midrule
\bmSection{} & \textbf{0.16}          \\
\bmAll{}     & 0.14 \\
\bmTitle{}   & 0.10          \\
\bottomrule
\end{tabular}
\hfill{}
\begin{tabular}{ll}
\toprule
Clustering Method       & ARI             \\ \midrule
\qqqsmMean{}    & \textbf{0.30} \\
\qqqsmLead{}    & 0.30          \\
\qqqsmTitle{}   & 0.29          \\
\sbertEuclid{} & 0.26          \\
\sbertCosine{}  & 0.26          \\ \bottomrule
\end{tabular}

\begin{comment}
\begin{tabular}{ll}
\toprule
Retrieval Method       & MAP             \\ \midrule
\bmSection{} & \textbf{0.1566}          \\
\bmAll{}     & 0.1429 \\
\bmTitle{}   & 0.1010          \\
\bottomrule
\end{tabular}
\hfill{}
\begin{tabular}{ll}
\toprule
Clustering Method       & ARI             \\ \midrule
\qqqsmMean{}    & \textbf{0.3002} \\
\qqqsmLead{}    & 0.2983          \\
\qqqsmTitle{}   & 0.2891          \\
\sbertEuclid{} & 0.2631          \\
\sbertCosine{}  & 0.2585          \\ \bottomrule
\end{tabular}
\end{comment}

\end{footnotesize}

\label{tab:res_comp}
\end{table}

\begin{table}
\caption{Comparison of different Summarizers on ``manual''.\label{tab:summarizers}}
\begin{footnotesize}

\begin{tabular}{llllr@{\extracolsep{0pt}$\pm$}l}
\toprule 
Summarization Method  & ROUGE-1 & ROUGE-2 & BERTScore & \multicolumn{2}{c}{Avg Length}\tabularnewline
\midrule 
\TLouvain{} short & 0.172 & 0.028 & 0.813 & \multicolumn{2}{r}{458\ld{$\pm$ 26}}\tabularnewline
\TLouvain{} long & 0.199 & \textbf{0.064} & 0.808 & \multicolumn{2}{r}{856}\tabularnewline
\TRedundancy{} & 0.179 & 0.035 & 0.812 & \multicolumn{2}{r}{1041\ld{$\pm$75}}\tabularnewline
Hier.\ Transformer \cite{liu2019hierarchical} & 0.074 & 0.013 & 0.770 & \multicolumn{2}{r}{76 \ld{$\pm$9}}\tabularnewline
BART-Long \cite{pasunuru2021efficiently}  & \textbf{0.211} & 0.052 & \textbf{0.818} & \multicolumn{2}{r}{827\ld{$\pm$19}}\tabularnewline
\bottomrule
\end{tabular}
\end{footnotesize}
\end{table}

\begin{table*}[t]
\caption{Results for systems with \TLouvain{} on the TREC-CAR Benchmark Y1 Test dataset. Bold entries indicate the highest measure in that column. \up and \down indicate significant change for $p < 0.05$ compared to the component-wise best (marked with \base). $\dagger$ marks methods that have access to oracle knowledge about subtopics and query descriptions, which are known to improve quality.
 \label{tab:overall}}
%\bmSection{} / \qqqsmMean{} combination

\begin{footnotesize}
% Preview source code for paragraph 0

\begin{tabular}{@{}llrrrrrrrrr@{}}
\toprule 
\multicolumn{1}{c}{Retrieval} & \multicolumn{1}{c}{Clustering} & \multicolumn{3}{c}{ROUGE-1} & \multicolumn{3}{c}{ROUGE-2} & \multicolumn{3}{c}{BERTScore}\tabularnewline
\cmidrule(lr) {3-5} \cmidrule(lr){6-8} \cmidrule(lr){9-11}  &  & \multicolumn{1}{c}{Precision} & \multicolumn{1}{c}{Recall} & \multicolumn{1}{c}{F-1} & \multicolumn{1}{c}{Precision} & \multicolumn{1}{c}{Recall} & \multicolumn{1}{c}{F-1} & \multicolumn{1}{c}{Precision} & \multicolumn{1}{c}{Recall} & \multicolumn{1}{c}{F-1}\tabularnewline
\midrule 
\multirow{5}{*}{ \bmSection{} $\dagger$ \base } & \qqqsmMean{}\base  & \base 0.525  & \base 0.097  & \base 0.151  & \base 0.238  & \base \textbf{0.033}  & \base \textbf{0.053}  & \base \textbf{0.815}  & \base 0.809  & \base\textbf{0.812}\tabularnewline
 & \qqqsmLead{} $\dagger$  & \up 0.535  & \down 0.087  & \down 0.139  & 0.241  & 0.029  & \down 0.047  & \down 0.814  & 0.811  & \textbf{0.812} \tabularnewline
 & \qqqsmTitle{}  & 0.530  & \down 0.087  & \down 0.138  & 0.238  & \down 0.028  & \down 0.046  & \down 0.813  & 0.810  & 0.811 \tabularnewline
 & \sbertCosine{}  & 0.516  & \down 0.092  & \down 0.144  & \down 0.227  & \down 0.029  & \down 0.047  & 0.814  & 0.809  & 0.811 \tabularnewline
 & \sbertEuclid{}  & 0.529  & \down 0.084  & \down 0.136  & 0.236  & \down 0.026  & \down 0.044  & \down 0.813  & 0.811  & \textbf{0.812} \tabularnewline
\midrule 
\multirow{5}{*}{\bmAll{} $\dagger$} & \qqqsmMean{}  & 0.537  & \textbf{0.099}  & \textbf{0.156}  & 0.235  & 0.032  & 0.052  & 0.814  & 0.810  & \textbf{0.812} \tabularnewline
 & \qqqsmLead{} $\dagger$  & \up 0.533  & 0.093  & 0.148  & \textbf{0.247}  & 0.031  & 0.051  & 0.813  & 0.811  & \textbf{0.812} \tabularnewline
 & \qqqsmTitle{}  & \up \textbf{0.548}  & 0.090  & 0.144  & 0.241  & 0.029  & 0.048  & 0.812  & 0.811  & \textbf{0.812} \tabularnewline
 & \sbertCosine{}  & 0.530  & 0.093  & 0.148  & 0.229  & 0.030  & 0.049  & 0.813  & 0.809  & 0.811 \tabularnewline
 & \sbertEuclid{}  & \up 0.541  & \down 0.089  & 0.144  & 0.235  & \down 0.028  & \down 0.046  & \down 0.812  & 0.810  & 0.811 \tabularnewline
\midrule 
\multirow{5}{*}{\bmTitle{}} & \qqqsmMean{}  & 0.512  & \down 0.082  & \down 0.127  & \down 0.220  & 0.030  & \down 0.046  & \down 0.809  & 0.811  & 0.810 \tabularnewline
 & \qqqsmLead{} $\dagger$  & 0.519  & \down 0.069  & \down 0.111  & 0.223  & \down 0.023  & \down 0.038  & \down 0.806  & \up 0.812  & \down 0.809 \tabularnewline
 & \qqqsmTitle{}  & 0.523  & \down 0.071  & \down 0.114  & 0.228  & \down 0.024  & \down 0.039  & \down 0.807  & \up \textbf{0.813}  & \down 0.810 \tabularnewline
 & \sbertCosine{}  & \down 0.494  & \down 0.072  & \down 0.114  & \down 0.206  & \down 0.024  & \down 0.038  & \down 0.807  & 0.810  & \down 0.808 \tabularnewline
 & \sbertEuclid{}  & \down 0.507  & \down 0.067  & \down 0.108  & \down 0.214  & \down 0.022  & \down 0.036  & \down 0.805  & 0.811  & \down 0.808 \tabularnewline
\midrule 
Manual $\dagger$  & Manual $\dagger$  & \down 0.409  & \up 0.144  & \up 0.199  & \down 0.162  & \up 0.045  & \up 0.064  & 0.814  & \down 0.802  & \down 0.808 \tabularnewline
\bottomrule
\end{tabular}

\end{footnotesize}

\end{table*}

% \vspace{-1em}
\subsection{Evaluation Metrics}

To evaluate the quality of the generated articles, we compare them to the content of the true Wikipedia page in the dataset. We lexical ROUGE-1 and ROUGE-2 \cite{lin2004rouge} and semantic BERTScore \cite{zhang2019bertscore}. %These metrics enable us to study both the linguistic and semantic overlap.

\cut{ROUGE-based metrics compare the linguistic overlap between a generated article and the ground truth via n-gram matching. BERTScore compares semantic representations of text by first embedding the tokens of each article with a pre-trained contextual embedding model before finding an alignment of matching tokens and comparing the maximum possible similarity.}

For the component-wise evaluation, we use (Mean) Average Precision (MAP) to evaluate retrieval and the Adjusted Rand Index (ARI) to evaluate clustering."

% \vspace{-1em}
%\subsection{Manual Baseline and Compared Systems}

\subsection{Compared Systems}

We compare different variations of our Retrieve-Cluster-Summarize system to a range of external systems and manual methods.

\paragraph{Manual} We simulate a perfect retrieval and clustering systems, using manual section-level assessments created by NIST for the TREC CAR track: All paragraphs that are assessed as relevant for the same Wikipedia section are taken as one cluster. In cases where the same paragraph is judged relevant for multiple sections, we prefer higher relevance assessments and randomly break ties. All clusters are summarized and composed into an article.

We compare to external summarization systems which use their own internal ranking/clustering modules:

\paragraph{Hierarchical Transformer \cite{liu2019hierarchical}} An abstractive summarization model that uses its own internal ranking module and exploits inter-document connections via the attention mechanism.

\paragraph{BART-Long \cite{pasunuru2021efficiently}} An extractive summarizer which uses the pre-trained BART model select content tokens. The summarizer incorporates a graph from Open-IE relation extractions to identify clusters for consolidation (Note: these clusters are different from the topical clusters we obtain in the clustering stage).

\progress{End of page 3}

\section{Results} \label{sec:results}

\paragraph{Performance of summarizers}
In Table \ref{tab:summarizers} we compare different summarization approaches. We see that \TLouvain{} obtains reasonable performance when compared to more complex summarizers.
The system Hierarchical transformer  \cite{liu2019hierarchical} is unable to compete with other systems, as it produces summaries that are too short. The system BART-Long \cite{pasunuru2021efficiently} is a very strong system, and we see that our system is able to perform similarly. 
 %
%Different summarization systems produce articles of vastly different length, which affects their performance on recall-oriented summarization measures such as ROUGE and BertScore.  The Hierarchical transformer obtains very short summaries for which it is penalized. %We believe it is worthwhile to explore this space in future work, using our baseline system and evaluation paradigm.
%
For \textit{\TLouvain{} long}, we adjust the aggressiveness of the Louvain redundancy removal component to produce articles that are about the same length as BART-Long and use it in the remainder of this experiment. We obtain nearly the same performance as the BART-Long system. Hence, we see that our piece-wise trained system is competitive to BART-Long.

\paragraph{Performance of individual components}
The results in Table \ref{tab:res_comp} display the individual performance of each retrieval and clustering method.
They show that both topic-enhanced models \bmSection{} followed by \bmAll{} are performing better than  \bmTitle{}. 

For clustering, the best method is \qqqsmMean{}, with other \qqqsm{} methods outperforming Sentence-BERT.

\paragraph{Overall system performance} Table \ref{tab:overall} compares all combinations of compount systems with their overall system evaluation. 
Despite each components being only trained on their respective sub-task benchmarks, the system comprised of the best components (first row) either obtains the best system-wise performance. (For ROUGE-1 it is not the best system, but it is not significantly outperformed by any other combination) Hence, we suggest this combination as good baseline for further research (\rqGoodBaselineSystem{}).

We find that \bmTitle{} yields the lowest overall performance. The query-agnostic Sentence-BERT leads to generally lower performance than the query-aware clustering method \qqqsm{}. These findings are in line with prior work \cite{kashyapi2022query,dietz2017trec}.
This supports our hypothesis that our evaluation paradigm allows for component-wise training (\rqWikimarksEvaluation{}).
%
%
%%We observe some general trends when comparing various system combinations to one another. We see the best average performance when the \bmSection{} retrieval method is used, regardless of clustering method. This is followed by the \bmAll{} retrieval, and finally the \bmTitle{}. We identify similar relationships between Clustering methods: using \qqqsmMean{} clustering nearly always results in the best summarization performance across all studied retrieval methods. 
%
%%Across linguistic-overlap metrics (ROUGE-based), we see a decrease in F-1 performance when comparing the system to the manual method, which simulates retrieval and clusters with manual relevance data from NIST assessors. 
%
%
%The articles produced with manual clusters have an average length of 856 words, which is  longer than those created by the joint system (average of 313 words). This leads to higher recall but lower precision, which drives the change in F-1 performance.
%
Regarding \rqManualSummarization{}, the retrieval and clustering system leads to overall performance that is competitive with manual assessments.

%% This result is driven primarily by a significantly higher recall when using manually retrieved documents. However, the manual input provides significantly lower precision when compared to all realistic inputs.

%% This may be due to differences in generated article length: articles produced via the manually created inputs have an average length of 856 words, while all articles created by our system have an average length of 313 words.

%%\ld{What?} The significant decline in F-1 performance with non-manual retrieval and clustering addresses R?? (impacts of non-ideal inputs) by indicating a clear relationship between upstream component performance and downstream results. Regarding \rqWikimarksEvaluation{} (correlations between components), we observe that better individual component performance directly translates to better overall system quality. Finally, regarding \rqGoodBaselineSystem{} (best set of upstream components), we identify the article generation system to be that which uses \bmSection{} and \qqqsmMean{}. This demonstrates the benefit of using a query-specific clustering system.

\progress{3/4 down page 4}

\section{Conclusion}
\label{sec:conclusion}

In this early work we study joint retrieval, clustering, and summarization systems for generating query-specific articles. We find that the best system for joint retrieval, clustering, and summarization is the one that uses \bmSection{} for retrieval, followed by query-specific clustering using \qqqsmMean{}. This system achieves the best per-component performance and is not outperformed by any other combination. 

We demonstrate strong positive correlation between the performance of individual components and the quality of resulting articles, even though each component is only trained on its own benchmark. These results demonstrate the effectiveness of the Wikimarks approach for generating benchmarks to train different components that work together seamlessly.

% Here we focused on developing a baseline system and an evaluation paradigm for integrating retrieval, clustering, and summarization to retrieve \emph{information} rather than just documents. This lays the groundwork for future research on optimizing the quality of generated query-specific articles by improving the  retrieval system.
% In future work, we plan to explore a wider range of neural retrieval models and investigate how they can be integrated with subtopic detection and natural language generation.

%%When comparing manually retrieved and clustered documents with our best performing system, we see increases in linguistic overlap scores and minor decrease in semantic overlap scores. 
% We note that the usage of manually simulated retrieval and clustering generally results in higher recall but lower precision for the linguistic overlap metrics on which we test. 

%%Care should be taken when designing systems that rely on upstream tasks such as retrieval and clustering. As demonstrated here, evaluation on components in isolation is a good indicator of joint performance, however it is critical to evaluate systems in their entirety.

%%These results provide important context when considering the usage of summarization models within a larger system. While it is standard to evaluate summarization models using idealized inputs and carefully constructed datasets, we must also consider the potential degradation in performance when dependencies on other components are present, such as in retrieval and clustering.

We empirically demonstrate  that the Wikimarks approach allows us to create different benchmarks for training individual components, so that these are work together synergistically to obtain ``the best of all worlds``. Thanks to this benchmark creation approach, we are able to overcome the dreaded error-propagation of systems trained in a pipelined fashion which usually leads to a degradation in overall-system performance. Instead with this approach we can train/select components individually and empirically demonstrate that these provide best system performance. Hence, when end-to-end training is too expensive, our benchmark creation approach offers a feasible alternative. If desired, our training approach can be further refined with a few epochs of end-to-end training. 

    \subsection*{Acknowledgements}

    This material is based upon work supported by the National Science Foundation under
    Grant No. 1846017. Any opinions, findings, and conclusions or recommendations expressed in this material
    are those of the author(s) and do not necessarily reflect the views of the National Science
    Foundation.

\bibliographystyle{ACM-Reference-Format}
\balance
\bibliography{bibliography}


\begin{thebibliography}{32}

%%% the Web address.

\ifx \showCODEN    \undefined \def \showCODEN     #1{\unskip}     \fi
\ifx \showDOI      \undefined \def \showDOI       #1{#1}\fi
\ifx \showISBNx    \undefined \def \showISBNx     #1{\unskip}     \fi
\ifx \showISBNxiii \undefined \def \showISBNxiii  #1{\unskip}     \fi
\ifx \showISSN     \undefined \def \showISSN      #1{\unskip}     \fi
\ifx \showLCCN     \undefined \def \showLCCN      #1{\unskip}     \fi
\ifx \shownote     \undefined \def \shownote      #1{#1}          \fi
\ifx \showarticletitle \undefined \def \showarticletitle #1{#1}   \fi
\ifx \showURL      \undefined \def \showURL       {\relax}        \fi
% invisible to TeX
\providecommand\bibfield[2]{#2}
\providecommand\bibinfo[2]{#2}
\providecommand\natexlab[1]{#1}
\providecommand\showeprint[2][]{arXiv:#2}

\bibitem[Allan et~al\mbox{.}(2012)]%
        {allan2012frontiers}
\bibfield{author}{\bibinfo{person}{James Allan}, \bibinfo{person}{Bruce Croft},
  \bibinfo{person}{Alistair Moffat}, {and} \bibinfo{person}{Mark Sanderson}.}
  \bibinfo{year}{2012}\natexlab{}.
\newblock \showarticletitle{Frontiers, challenges, and opportunities for
  information retrieval: Report from SWIRL 2012 the second strategic workshop
  on information retrieval in Lorne}. In \bibinfo{booktitle}{\emph{Acm sigir
  forum}}, Vol.~\bibinfo{volume}{46}. ACM New York, NY, USA,
  \bibinfo{pages}{2--32}.
\newblock


\bibitem[Banerjee and Mitra(2015)]%
        {banerjee2015wikikreator}
\bibfield{author}{\bibinfo{person}{Siddhartha Banerjee} {and}
  \bibinfo{person}{Prasenjit Mitra}.} \bibinfo{year}{2015}\natexlab{}.
\newblock \showarticletitle{{W}iki{K}reator: Improving {W}ikipedia Stubs
  Automatically}. In \bibinfo{booktitle}{\emph{Proceedings of the 53rd Annual
  Meeting of the Association for Computational Linguistics and the 7th
  International Joint Conference on Natural Language Processing (Volume 1: Long
  Papers)}}. \bibinfo{publisher}{Association for Computational Linguistics},
  \bibinfo{address}{Beijing, China}, \bibinfo{pages}{867--877}.
\newblock
\urldef\tempurl%
\url{https://doi.org/10.3115/v1/P15-1084}
\showDOI{\tempurl}


\bibitem[Bernardini et~al\mbox{.}(2009)]%
        {bernardini2009full}
\bibfield{author}{\bibinfo{person}{Andrea Bernardini}, \bibinfo{person}{Claudio
  Carpineto}, {and} \bibinfo{person}{Massimiliano D'Amico}.}
  \bibinfo{year}{2009}\natexlab{}.
\newblock \showarticletitle{Full-subtopic retrieval with keyphrase-based search
  results clustering}. In \bibinfo{booktitle}{\emph{2009 IEEE/WIC/ACM
  International Joint Conference on Web Intelligence and Intelligent Agent
  Technology}}, Vol.~\bibinfo{volume}{1}. IEEE, \bibinfo{pages}{206--213}.
\newblock


\bibitem[Blei et~al\mbox{.}(2003)]%
        {blei2003latent}
\bibfield{author}{\bibinfo{person}{David~M Blei}, \bibinfo{person}{Andrew~Y
  Ng}, {and} \bibinfo{person}{Michael~I Jordan}.}
  \bibinfo{year}{2003}\natexlab{}.
\newblock \showarticletitle{Latent dirichlet allocation}.
\newblock \bibinfo{journal}{\emph{Journal of machine Learning research}}
  \bibinfo{volume}{3}, \bibinfo{number}{Jan} (\bibinfo{year}{2003}),
  \bibinfo{pages}{993--1022}.
\newblock


\bibitem[Blondel et~al\mbox{.}(2008)]%
        {blondel2008fast}
\bibfield{author}{\bibinfo{person}{Vincent~D Blondel},
  \bibinfo{person}{Jean-Loup Guillaume}, \bibinfo{person}{Renaud Lambiotte},
  {and} \bibinfo{person}{Etienne Lefebvre}.} \bibinfo{year}{2008}\natexlab{}.
\newblock \showarticletitle{Fast unfolding of communities in large networks}.
\newblock \bibinfo{journal}{\emph{Journal of statistical mechanics: theory and
  experiment}} \bibinfo{volume}{2008}, \bibinfo{number}{10}
  (\bibinfo{year}{2008}), \bibinfo{pages}{P10008}.
\newblock


\bibitem[Carpineto and Romano(2012a)]%
        {carpineto2012consensus}
\bibfield{author}{\bibinfo{person}{Claudio Carpineto} {and}
  \bibinfo{person}{Giovanni Romano}.} \bibinfo{year}{2012}\natexlab{a}.
\newblock \showarticletitle{Consensus clustering based on a new probabilistic
  rand index with application to subtopic retrieval}.
\newblock \bibinfo{journal}{\emph{IEEE Transactions on Pattern Analysis and
  Machine Intelligence}} \bibinfo{volume}{34}, \bibinfo{number}{12}
  (\bibinfo{year}{2012}), \bibinfo{pages}{2315--2326}.
\newblock


\bibitem[Carpineto and Romano(2012b)]%
        {carpineto2012survey}
\bibfield{author}{\bibinfo{person}{Claudio Carpineto} {and}
  \bibinfo{person}{Giovanni Romano}.} \bibinfo{year}{2012}\natexlab{b}.
\newblock \showarticletitle{A survey of automatic query expansion in
  information retrieval}.
\newblock \bibinfo{journal}{\emph{Acm Computing Surveys (CSUR)}}
  \bibinfo{volume}{44}, \bibinfo{number}{1} (\bibinfo{year}{2012}),
  \bibinfo{pages}{1--50}.
\newblock


\bibitem[Claveau(2021)]%
        {claveau2021neural}
\bibfield{author}{\bibinfo{person}{Vincent Claveau}.}
  \bibinfo{year}{2021}\natexlab{}.
\newblock \showarticletitle{Neural text generation for query expansion in
  information retrieval}. In \bibinfo{booktitle}{\emph{IEEE/WIC/ACM
  International Conference on Web Intelligence and Intelligent Agent
  Technology}}. \bibinfo{pages}{202--209}.
\newblock


\bibitem[Dietz et~al\mbox{.}(2022)]%
        {dietz2022wikimarks}
\bibfield{author}{\bibinfo{person}{Laura Dietz}, \bibinfo{person}{Shubham
  Chatterjee}, \bibinfo{person}{Connor Lennox}, \bibinfo{person}{Sumanta
  Kashyapi}, \bibinfo{person}{Pooja Oza}, {and} \bibinfo{person}{Ben Gamari}.}
  \bibinfo{year}{2022}\natexlab{}.
\newblock \showarticletitle{Wikimarks: Harvesting Relevance Benchmarks from
  Wikipedia}. In \bibinfo{booktitle}{\emph{Proceedings of the 45th
  International ACM SIGIR Conference on Research and Development in Information
  Retrieval}}. \bibinfo{pages}{3003--3012}.
\newblock


\bibitem[Dietz and Dalton(2020)]%
        {dietz2020humans}
\bibfield{author}{\bibinfo{person}{Laura Dietz} {and} \bibinfo{person}{Jeff
  Dalton}.} \bibinfo{year}{2020}\natexlab{}.
\newblock \showarticletitle{Humans optional? automatic large-scale test
  collections for entity, passage, and entity-passage retrieval}.
\newblock \bibinfo{journal}{\emph{Datenbank-Spektrum}} \bibinfo{volume}{20},
  \bibinfo{number}{1} (\bibinfo{year}{2020}), \bibinfo{pages}{17--28}.
\newblock


\bibitem[Dietz et~al\mbox{.}(2017)]%
        {dietz2017trec}
\bibfield{author}{\bibinfo{person}{Laura Dietz}, \bibinfo{person}{Manisha
  Verma}, \bibinfo{person}{Filip Radlinski}, {and} \bibinfo{person}{Nick
  Craswell}.} \bibinfo{year}{2017}\natexlab{}.
\newblock \showarticletitle{TREC Complex Answer Retrieval Overview.}. In
  \bibinfo{booktitle}{\emph{TREC}}.
\newblock


\bibitem[Guo et~al\mbox{.}(2023)]%
        {guo2023close}
\bibfield{author}{\bibinfo{person}{Biyang Guo}, \bibinfo{person}{Xin Zhang},
  \bibinfo{person}{Ziyuan Wang}, \bibinfo{person}{Minqi Jiang},
  \bibinfo{person}{Jinran Nie}, \bibinfo{person}{Yuxuan Ding},
  \bibinfo{person}{Jianwei Yue}, {and} \bibinfo{person}{Yupeng Wu}.}
  \bibinfo{year}{2023}\natexlab{}.
\newblock \showarticletitle{How Close is ChatGPT to Human Experts? Comparison
  Corpus, Evaluation, and Detection}.
\newblock \bibinfo{journal}{\emph{arXiv preprint arXiv:2301.07597}}
  (\bibinfo{year}{2023}).
\newblock


\bibitem[Hayashi et~al\mbox{.}(2021)]%
        {hayashi2021wikiasp}
\bibfield{author}{\bibinfo{person}{Hiroaki Hayashi}, \bibinfo{person}{Prashant
  Budania}, \bibinfo{person}{Peng Wang}, \bibinfo{person}{Chris Ackerson},
  \bibinfo{person}{Raj Neervannan}, {and} \bibinfo{person}{Graham Neubig}.}
  \bibinfo{year}{2021}\natexlab{}.
\newblock \showarticletitle{WikiAsp: A Dataset for Multi-domain Aspect-based
  Summarization}.
\newblock \bibinfo{journal}{\emph{Transactions of the Association for
  Computational Linguistics}}  \bibinfo{volume}{9} (\bibinfo{year}{2021}),
  \bibinfo{pages}{211--225}.
\newblock


\bibitem[Izacard et~al\mbox{.}(2022)]%
        {izacard2022atlas}
\bibfield{author}{\bibinfo{person}{Gautier Izacard}, \bibinfo{person}{Patrick
  Lewis}, \bibinfo{person}{Maria Lomeli}, \bibinfo{person}{Lucas Hosseini},
  \bibinfo{person}{Fabio Petroni}, \bibinfo{person}{Timo Schick},
  \bibinfo{person}{Jane Dwivedi-Yu}, \bibinfo{person}{Armand Joulin},
  \bibinfo{person}{Sebastian Riedel}, {and} \bibinfo{person}{Edouard Grave}.}
  \bibinfo{year}{2022}\natexlab{}.
\newblock \showarticletitle{Atlas: Few-shot learning with retrieval augmented
  language models}.
\newblock \bibinfo{journal}{\emph{arXiv preprint arXiv}}
  \bibinfo{volume}{2208} (\bibinfo{year}{2022}).
\newblock


\bibitem[Kashyapi and Dietz(2022)]%
        {kashyapi2022query}
\bibfield{author}{\bibinfo{person}{Sumanta Kashyapi} {and}
  \bibinfo{person}{Laura Dietz}.} \bibinfo{year}{2022}\natexlab{}.
\newblock \showarticletitle{Query-specific Subtopic Clustering}.
\newblock \bibinfo{journal}{\emph{ACM/IEEE Joint Conference on Digital
  Libraries (JCDL}} (\bibinfo{year}{2022}).
\newblock


\bibitem[Lewis et~al\mbox{.}(2020)]%
        {lewis2020retrieval}
\bibfield{author}{\bibinfo{person}{Patrick Lewis}, \bibinfo{person}{Ethan
  Perez}, \bibinfo{person}{Aleksandra Piktus}, \bibinfo{person}{Fabio Petroni},
  \bibinfo{person}{Vladimir Karpukhin}, \bibinfo{person}{Naman Goyal},
  \bibinfo{person}{Heinrich K{\"u}ttler}, \bibinfo{person}{Mike Lewis},
  \bibinfo{person}{Wen-tau Yih}, \bibinfo{person}{Tim Rockt{\"a}schel},
  {et~al\mbox{.}}} \bibinfo{year}{2020}\natexlab{}.
\newblock \showarticletitle{Retrieval-augmented generation for
  knowledge-intensive nlp tasks}.
\newblock \bibinfo{journal}{\emph{Advances in Neural Information Processing
  Systems}}  \bibinfo{volume}{33} (\bibinfo{year}{2020}),
  \bibinfo{pages}{9459--9474}.
\newblock


\bibitem[Li et~al\mbox{.}(2020)]%
        {li2020leveraging}
\bibfield{author}{\bibinfo{person}{Wei Li}, \bibinfo{person}{Xinyan Xiao},
  \bibinfo{person}{Jiachen Liu}, \bibinfo{person}{Hua Wu},
  \bibinfo{person}{Haifeng Wang}, {and} \bibinfo{person}{Junping Du}.}
  \bibinfo{year}{2020}\natexlab{}.
\newblock \showarticletitle{Leveraging Graph to Improve Abstractive
  Multi-Document Summarization}. In \bibinfo{booktitle}{\emph{Proceedings of
  the 58th Annual Meeting of the Association for Computational Linguistics}}.
  \bibinfo{pages}{6232--6243}.
\newblock


\bibitem[Lin(2004)]%
        {lin2004rouge}
\bibfield{author}{\bibinfo{person}{Chin-Yew Lin}.}
  \bibinfo{year}{2004}\natexlab{}.
\newblock \showarticletitle{Rouge: A package for automatic evaluation of
  summaries}. In \bibinfo{booktitle}{\emph{Text summarization branches out}}.
  \bibinfo{pages}{74--81}.
\newblock


\bibitem[Lin and Hovy(2002)]%
        {lin2002single}
\bibfield{author}{\bibinfo{person}{Chin-Yew Lin} {and} \bibinfo{person}{Eduard
  Hovy}.} \bibinfo{year}{2002}\natexlab{}.
\newblock \showarticletitle{From single to multi-document summarization}. In
  \bibinfo{booktitle}{\emph{Proceedings of the 40th annual meeting of the
  association for computational linguistics}}. \bibinfo{pages}{457--464}.
\newblock


\bibitem[Liu et~al\mbox{.}(2018)]%
        {liu2018generating}
\bibfield{author}{\bibinfo{person}{Peter~J Liu}, \bibinfo{person}{Mohammad
  Saleh}, \bibinfo{person}{Etienne Pot}, \bibinfo{person}{Ben Goodrich},
  \bibinfo{person}{Ryan Sepassi}, \bibinfo{person}{Lukasz Kaiser}, {and}
  \bibinfo{person}{Noam Shazeer}.} \bibinfo{year}{2018}\natexlab{}.
\newblock \showarticletitle{Generating wikipedia by summarizing long
  sequences}.
\newblock \bibinfo{journal}{\emph{arXiv preprint arXiv:1801.10198}}
  (\bibinfo{year}{2018}).
\newblock


\bibitem[Liu and Lapata(2019)]%
        {liu2019hierarchical}
\bibfield{author}{\bibinfo{person}{Yang Liu} {and} \bibinfo{person}{Mirella
  Lapata}.} \bibinfo{year}{2019}\natexlab{}.
\newblock \showarticletitle{Hierarchical Transformers for Multi-Document
  Summarization}. In \bibinfo{booktitle}{\emph{Proceedings of the 57th Annual
  Meeting of the Association for Computational Linguistics}}.
  \bibinfo{pages}{5070--5081}.
\newblock


\bibitem[Ma et~al\mbox{.}(2020)]%
        {ma2020multi}
\bibfield{author}{\bibinfo{person}{Congbo Ma}, \bibinfo{person}{Wei~Emma
  Zhang}, \bibinfo{person}{Mingyu Guo}, \bibinfo{person}{Hu Wang}, {and}
  \bibinfo{person}{Quan~Z Sheng}.} \bibinfo{year}{2020}\natexlab{}.
\newblock \showarticletitle{Multi-document summarization via deep learning
  techniques: A survey}.
\newblock \bibinfo{journal}{\emph{arXiv preprint arXiv:2011.04843}}
  (\bibinfo{year}{2020}).
\newblock


\bibitem[Ouyang et~al\mbox{.}(2022)]%
        {ouyangtraining}
\bibfield{author}{\bibinfo{person}{Long Ouyang}, \bibinfo{person}{Jeffrey Wu},
  \bibinfo{person}{Xu Jiang}, \bibinfo{person}{Diogo Almeida},
  \bibinfo{person}{Carroll Wainwright}, \bibinfo{person}{Pamela Mishkin},
  \bibinfo{person}{Chong Zhang}, \bibinfo{person}{Sandhini Agarwal},
  \bibinfo{person}{Katarina Slama}, \bibinfo{person}{Alex Gray},
  {et~al\mbox{.}}} \bibinfo{year}{2022}\natexlab{}.
\newblock \showarticletitle{Training language models to follow instructions
  with human feedback}. In \bibinfo{booktitle}{\emph{Advances in Neural
  Information Processing Systems}}.
\newblock


\bibitem[Pasunuru et~al\mbox{.}(2021)]%
        {pasunuru2021efficiently}
\bibfield{author}{\bibinfo{person}{Ramakanth Pasunuru},
  \bibinfo{person}{Mengwen Liu}, \bibinfo{person}{Mohit Bansal},
  \bibinfo{person}{Sujith Ravi}, {and} \bibinfo{person}{Markus Dreyer}.}
  \bibinfo{year}{2021}\natexlab{}.
\newblock \showarticletitle{Efficiently summarizing text and graph encodings of
  multi-document clusters}. In \bibinfo{booktitle}{\emph{Proceedings of the
  2021 Conference of the North American Chapter of the Association for
  Computational Linguistics: Human Language Technologies}}.
  \bibinfo{pages}{4768--4779}.
\newblock


\bibitem[Pradeep et~al\mbox{.}(2021)]%
        {pradeep2021expando}
\bibfield{author}{\bibinfo{person}{Ronak Pradeep}, \bibinfo{person}{Rodrigo
  Nogueira}, {and} \bibinfo{person}{Jimmy Lin}.}
  \bibinfo{year}{2021}\natexlab{}.
\newblock \showarticletitle{The expando-mono-duo design pattern for text
  ranking with pretrained sequence-to-sequence models}.
\newblock \bibinfo{journal}{\emph{arXiv preprint arXiv:2101.05667}}
  (\bibinfo{year}{2021}).
\newblock


\bibitem[Radford et~al\mbox{.}(2019)]%
        {radford2019language}
\bibfield{author}{\bibinfo{person}{Alec Radford}, \bibinfo{person}{Jeffrey Wu},
  \bibinfo{person}{Rewon Child}, \bibinfo{person}{David Luan},
  \bibinfo{person}{Dario Amodei}, \bibinfo{person}{Ilya Sutskever},
  {et~al\mbox{.}}} \bibinfo{year}{2019}\natexlab{}.
\newblock \showarticletitle{Language models are unsupervised multitask
  learners}.
\newblock \bibinfo{journal}{\emph{OpenAI blog}} \bibinfo{volume}{1},
  \bibinfo{number}{8} (\bibinfo{year}{2019}), \bibinfo{pages}{9}.
\newblock


\bibitem[Raffel et~al\mbox{.}(2019)]%
        {raffel2019exploring}
\bibfield{author}{\bibinfo{person}{Colin Raffel}, \bibinfo{person}{Noam
  Shazeer}, \bibinfo{person}{Adam Roberts}, \bibinfo{person}{Katherine Lee},
  \bibinfo{person}{Sharan Narang}, \bibinfo{person}{Michael Matena},
  \bibinfo{person}{Yanqi Zhou}, \bibinfo{person}{Wei Li}, {and}
  \bibinfo{person}{Peter~J Liu}.} \bibinfo{year}{2019}\natexlab{}.
\newblock \showarticletitle{Exploring the limits of transfer learning with a
  unified text-to-text transformer}.
\newblock \bibinfo{journal}{\emph{arXiv preprint arXiv:1910.10683}}
  (\bibinfo{year}{2019}).
\newblock


\bibitem[Raiber and Kurland(2013)]%
        {raiber2013ranking}
\bibfield{author}{\bibinfo{person}{Fiana Raiber} {and} \bibinfo{person}{Oren
  Kurland}.} \bibinfo{year}{2013}\natexlab{}.
\newblock \showarticletitle{Ranking document clusters using markov random
  fields}. In \bibinfo{booktitle}{\emph{Proceedings of the 36th international
  ACM SIGIR conference on Research and development in information retrieval}}.
  \bibinfo{pages}{333--342}.
\newblock


\bibitem[Reimers and Gurevych(2019)]%
        {reimers2019sentence}
\bibfield{author}{\bibinfo{person}{Nils Reimers} {and} \bibinfo{person}{Iryna
  Gurevych}.} \bibinfo{year}{2019}\natexlab{}.
\newblock \showarticletitle{Sentence-BERT: Sentence Embeddings using Siamese
  BERT-Networks}. In \bibinfo{booktitle}{\emph{Proceedings of the 2019
  Conference on Empirical Methods in Natural Language Processing and the 9th
  International Joint Conference on Natural Language Processing
  (EMNLP-IJCNLP)}}. \bibinfo{pages}{3982--3992}.
\newblock


\bibitem[Santos et~al\mbox{.}(2015)]%
        {santos2015search}
\bibfield{author}{\bibinfo{person}{Rodrygo~LT Santos}, \bibinfo{person}{Craig
  Macdonald}, \bibinfo{person}{Iadh Ounis}, {et~al\mbox{.}}}
  \bibinfo{year}{2015}\natexlab{}.
\newblock \showarticletitle{Search result diversification}.
\newblock \bibinfo{journal}{\emph{Foundations and Trends{\textregistered} in
  Information Retrieval}} \bibinfo{volume}{9}, \bibinfo{number}{1}
  (\bibinfo{year}{2015}), \bibinfo{pages}{1--90}.
\newblock


\bibitem[Sauper and Barzilay(2009)]%
        {sauper2009automatically}
\bibfield{author}{\bibinfo{person}{Christina Sauper} {and}
  \bibinfo{person}{Regina Barzilay}.} \bibinfo{year}{2009}\natexlab{}.
\newblock \showarticletitle{Automatically Generating {W}ikipedia Articles: A
  Structure-Aware Approach}. In \bibinfo{booktitle}{\emph{Proceedings of the
  Joint Conference of the 47th Annual Meeting of the {ACL} and the 4th
  International Joint Conference on Natural Language Processing of the
  {AFNLP}}}. \bibinfo{publisher}{Association for Computational Linguistics},
  \bibinfo{address}{Suntec, Singapore}, \bibinfo{pages}{208--216}.
\newblock
\urldef\tempurl%
\url{https://aclanthology.org/P09-1024}
\showURL{%
\tempurl}


\bibitem[Zhang et~al\mbox{.}(2019)]%
        {zhang2019bertscore}
\bibfield{author}{\bibinfo{person}{Tianyi Zhang}, \bibinfo{person}{Varsha
  Kishore}, \bibinfo{person}{Felix Wu}, \bibinfo{person}{Kilian~Q Weinberger},
  {and} \bibinfo{person}{Yoav Artzi}.} \bibinfo{year}{2019}\natexlab{}.
\newblock \showarticletitle{Bertscore: Evaluating text generation with bert}.
\newblock \bibinfo{journal}{\emph{arXiv preprint arXiv:1904.09675}}
  (\bibinfo{year}{2019}).
\newblock


\end{thebibliography}

\end{document}